# Quantitatively Predicting Modal Thermal Conductivity of Nanocrystalline Si by *full band* Monte Carlo simulations

Lina Yang[1*], Yi Jiang[1], Yanguang Zhou[2*]

[1]School of Aerospace Engineering, Beijing Institute of Technology, Beijing 100081, China

[2]Department of Mechanical and Aerospace Engineering, The Hong Kong University of Science and Technology, Clear water bay, Kowloon, Hong Kong SAR, China

Email: yangln@bit.edu.cn (L. Yang), maeygzhou@ust.hk (Y. Zhou)

**Abstract**

Thermal transport of nanocrystalline Si is of great importance for the application of thermoelectrics. A better understanding of the modal thermal conductivity of nanocrystalline Si will be expected to benefit the efficiency of thermoelectrics. In this work, the variance reduced Monte Carlo simulation with full band of phonon dispersion is applied to study the modal thermal conductivity of nanocrystalline Si. Importantly, the phonon modal transmissions across the grain boundaries which are modeled by the amorphous Si interface are calculated by the mode-resolved atomistic Green's function method. The predicted ratios of thermal conductivity of nanocrystalline Si to that of bulk Si agree well with that of the experimental measurements in a wide range of grain size. The thermal conductivity of nanocrystalline Si is decreased from 54% to 3% and the contribution of phonons with mean free path larger than the grain size increases from 30% to 96% as the grain size decreases from 550 nm to 10 nm. This work demonstrates that the *full band* Monte Carlo simulation using phonon modal transmission by the mode-resolved atomistic Green's function method can capture the phonon transport picture in complex nanostructures, and therefore can provide guidance for designing high performance Si based thermoelectrics.

## 1. Introduction

Thermal transport properties of nanostructures are of great importance for the advanced applications including thermoelectrics[1-3] which can convert waste heat into electricity and work as solid-state refrigeration, microelectronics[4] and thermal barrier coatings[5,6]. Recent studies have shown that nanocrystalline materials can largely reduce the lattice thermal conductivity ($\kappa$), which can largely benefit the efficiency of thermoelectrics.[7-9] Silicon-based materials are relatively inexpensive and most are nontoxic in contrast to many other thermoelectric materials, which make them promising thermoelectrics.[10-14] Nanocrystalline Si (nc-Si) has been widely studied, it was found that nc-Si is competitive thermoelectric material with best ZT = 0.7 at 1275 K[15], but it is still lower than those of the champion thermoelectric materials primarily because of its relatively high thermal conductivity. Therefore, further understanding thermal transport in nanocrystalline structures is necessary.

In nanocrystalline materials, the remarkable reduction of thermal conductivity is caused by the impedance of phonons at the grain boundaries when the grain size approaches or is smaller than the phonon mean free path (MFP).[16-21] Therefore, grain size and phonon transmission at boundaries are two important factors to affect the thermal conductivity. Wang et al. investigated nc-Si with grain size varied from 550 nm to 76 nm[16], they found that thermal conductivities of nc-Si show a $T^2$ dependence at low temperature, a similar trend was also found in Si inverse opals[22], which cannot be explained by the traditional phonon gray model. They reported that the frequency dependent (nongray) model should be used for boundary scatterings[16]. Later, Jugdersuren et al. studied nanocrystalline Si with grain size decreased to ~10 nm[23]. The thermal conductivity of nanocrystalline Si with smaller grain size was also studied by molecular dynamic simulations[17,24], and it is found that the thermal conductivity was quickly decreased as the decrease of grain size (< 8 nm), which is caused by the restrain of phonon MFP by the nano-grain boundaries[17]. Although there are many works about the thermal transport of nanocrystalline materials, researches based on modal level analysis of thermal conductivity are few, which is critical for further understanding the underlying mechanisms of thermal transport of nanocrystalline materials.

To quantitatively investigate the modal contribution of phonon modes in nanocrystalline materials, modeling the phonon mode transport process in nanocrystals is required. Phonon Monte Carlo (MC) simulations have been used to study the thermal transport in many complex nanostructures including nanocrystallines.[25-29] Later, the traditional phonon MC method is speeded up on the order of $10^9$ by the variance-reduced Monte Carlo (VRMC) algorithm[30,31], which has been widely used in the study of thermal transport of complex structures under the assumption of isotropic phonon dispersion.[32,33] To obtain modal thermal conductivity, full band of phonon dispersion should be used in the calculations. Monte Carlo simulation with full band of phonon dispersion was applied in the study of thermal transport of bulk Si and Si structures[26,34,35], of which the simulation results are found to be close to the experiment measurements[34]. Importantly, it is reported that the correct implementation of the phonon dispersion relation in MC simulations is essential to accurately capture the quasi-ballistic phonon transport[35]. Full band MC simulation is also used to investigate the thermal transport of Si/Ge heterostructures, while the phonon transmissions across the interface are calculated using the diffusive mismatch model.[36] Later, the VRMC with full band of phonon dispersion using the optimized phonon transmission across the interfaces is used to study the thermal transport of nc-Si[37]. The calculated thermal conductivity of nc-Si is found to be quite close to the experimental measurements in a wide range of temperature. Apart from these recent progresses, using modal level interfacial phonon transmission in the study of thermal transport by VRMC with full band of phonon dispersion is important to provide a physical and reliable phonon transport picture in nanocrystals.

Many approaches have been used to calculate the phonon modal transmissions across the interfaces. For instance, the frequency dependent phonon transmission calculated by the spectral diffuse mismatch model (SDMM)[38] was integrated in the MC simulations to calculate the interfacial thermal conductance (ITC). Modal analysis of ITC is developed based on molecular dynamics which inherently includes anharmonicity effect, but it is hard to be applied to large systems due to its high computational requirement.[39,40] On the other hand, atomistic Green's function (AGF) method is more efficient and easier to implement, and therefore has been used extensively to compute the frequency dependent phonon transmission.[41-45] Several extended techniques based on the AGF method have been developed to compute phonon modal transmission.[46-49] A similar numerical method was developed using perfectly matched layer boundaries to compute modal transmission.[50] The scattering boundary method can be used to calculate the phonon modal transmission[51], which is theoretically equivalent to the AGF method. Recently, Ong and Zhang have extended the conventional AGF formalism to mode-resolved AGF, which can calculate phonon modal transmission.[48] The phonon modal transmission across the Si/Ge interface[52] and the amorphous Si (a-Si) interfaces[53] was investigated by the mode-resolved AGF method. For the amorphous Si interfaces, it was found that the interface acts as a low-pass filter, reflecting modes with frequency greater than around 3 THz while transmitting those below this frequency, which agrees with the experimental measurement[54]. Although phonon modal transmission by the mode-resolved AGF method has been used to calculate interfacial thermal conductance, few works have integrated it in the calculations of thermal transport in nanocrystals.

In this work, the modal thermal conductivity of nc-Si is investigated by the VRMC with full band of phonon dispersion using phonon modal transmission calculated by the mode-resolved AGF method. The grain boundaries are modeled by the amorphous Si interface. The thermal conductivity of nc-Si with grain size from 10 to 700 nm are calculated, and compared to the experiment works. The thickness of grain boundary which will affect the phonon modal transmission is modulated to study its effect on the thermal conductivity of nc-Si. Furthermore, the frequency and mean free path dependence of thermal conductivity are analyzed. In this work, the thermal conductivities of nc-Si are calculated at room temperature.

## 2. Model and method

A cubic simulation box oriented along the $x$, $y$ and $z$ directions as shown in Fig. 1 is used to simulate the nc-Si. A temperature gradient is applied in the $x$ direction to set up a heat flux. The periodic heat flux boundary condition[25] along $x$ direction and a specular boundary condition in the $y$ and $z$ directions are applied, so that the computational domain represents a unit cell which is repeated in all directions. Three grain boundaries presented by the red, blue and green planes in Fig. 1 are perpendicular to the $x$, $y$, and $z$ direction, respectively, and these grain boundaries bisect the simulation box. Due to the symmetry boundary conditions, the grain size is the same as that of the cubic simulation domain. The red dashed circle shows the zoom-in of the grain boundary, which is modeled by a-Si interface. The thickness of all the a-Si interfaces (red, blue and green planes in Fig. 1) is the same in the nc-Si. The thickness of a-Si interface can be modulated for different nc-Si models to manipulate the phonon modal transmission. The specularity in all the simulations are set as unit, since the a-Si interface is not severely rough and the phonon transmission have stronger effect on the thermal conductivity than specularity[37].

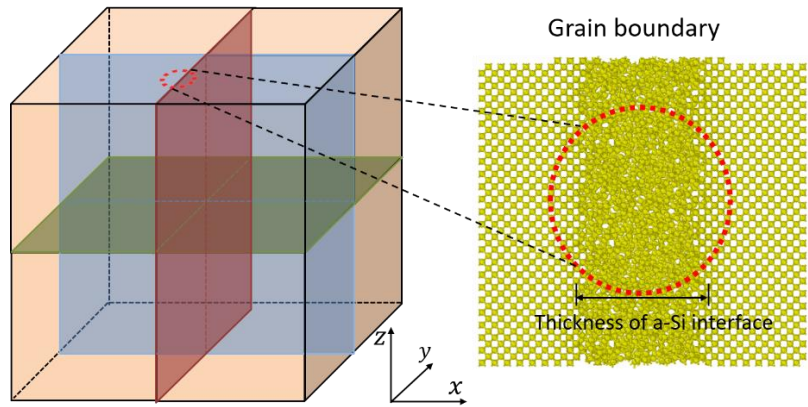

Fig. 1. The schematic of the nanocrystalline Si. The grain boundaries, which are represented by red, blue and green planes, are perpendicular to the $x$, $y$ and $z$ directions, respectively. The red dashed circle shows the zoom-in of the grain boundary, which is modeled by the amorphous Si interface. The thickness of all the a-Si interfaces (red, blue and green planes) is the same in the nc-Si. The thickness of a-Si interface can be modulated for different nc-Si models to manipulate the phonon modal transmission.

Further, the VRMC with full band of phonon dispersion is applied to simulate phonon transport in nc-Si (Fig. 1). The phonon transport is described by the Boltzmann transport equation under the relaxation time approximation. The deviational energy based Boltzmann transport equation[30] is given by

$$\frac{\partial e^d}{\partial t} + \boldsymbol{v_{k,p}} \cdot \nabla e^d = \frac{(e^{loc} - e^{eq}) - e^d}{\tau(\boldsymbol{k}, p, T)} \tag{1}$$

where $e^d = \hbar\omega(f - f_{T_{eq}}^{eq})$ is the desired deviational distribution function, $\boldsymbol{v_{k,p}}$ is the group velocity, $f_{T_{eq}}^{eq} = (\exp\left(\frac{\hbar\omega_{\boldsymbol{k,p}}}{k_B T_{eq}}\right) - 1)^{-1}$ is the Bose-Einstein distribution at the control temperature $T_{eq}$, $\omega_{\boldsymbol{k,p}}$ is the angular frequency, and $\tau(\boldsymbol{k}, p, T)$ is the relaxation time. Here $\boldsymbol{k}$ and $p$ denote the wave vector and polarization of a specific phonon mode, respectively. The Eq. (1) can be solved by using the linearized version of Peraud and Hadjiconstantinou's algorithm[31], and the details of using the VRMC with full band of phonon dispersion are in the Ref.[37].

The full band of phonon dispersion and phonon relaxation time of the three phonon scatterings, which can be calculated by ShengBTE package[55], are used as input for the Monte Carlo simulations. The conventional cell (CC) of bulk Si is used as unit cell ($1 \times 1 \times 1$ CC) for the calculation of phonon dispersion, which results in 24 polarizations. The interaction between Si atoms is described by the Tersoff potential[56]. The thermal conductivity of bulk Si is calculated as 239.5 W/m-K, which is consistent with the results of molecular dynamics simulation using the same potential[57,58]. The thermal conductivity calculated using Tersoff potential is overestimated comparing to the experimental measurements, while it is meaningful to investigate the trend and mechanism of the reduction of thermal conductivity in nanostructures. In the Monte Carlo simulation, 10,648 k points with 24 polarizations at each k are sampled over the entire Brillouin zone. The number of deviational particles is set as $N_{ph} = 8 \times 10^6$ for each simulation. $T_{eq}$ is set as 300 K, the temperature gradient and heat flux are along $x$ direction. The grain size can be controlled.

For the calculation of phonon modal transmission across the grain boundaries modeled using a-Si interface (Fig. 1), the mode-resolved AGF method[48] is applied. The key point for the calculation of phonon modal transmission is to construct the mode-resolved transmission matrix $\boldsymbol{t}$. The mode-resolved transmission for mode $i$ on one side of the interface coupled with mode $j$ on the other side of the interface is $\Xi_{ij} = \left| t_{ij} \right|^2$. The modal transmission of phonon mode $i$ can be obtained by summing over all possible phonon modes coupled with mode $i$, $\Xi_i = \sum_j \Xi_{ij}$. The spectral transmission $\Xi(\omega)$ is then calculated by summing over all the phonon modal transmission with angular frequency $\omega$, $\Xi(\omega) = \sum_i \Xi_i(\omega)$. In the following, the mode-resolved AGF is referred as AGF for simplicity. The phonon dispersion is calculated by lattice dynamics (LD) using the general utility lattice program (GULP).[59] The unit cell of the Si contacts for the calculation of phonon modal transmission is set as $1 \times 4 \times 4$ CC, which follows the settings in the Ref.[52], and the Tersoff potential[56] is used for depicting the interactions among atoms. The details of generating the a-Si interface, the settings of the devices and the calculation of phonon modal transmission can be found in Ref.[53]. The thickness of the a-Si interface can be modulated. The frequency interval in all the AGF calculation is set as 0.1 THz. Because the unit cell used in the AGF calculation is 4 times as large as that used in the MC simulations along $y$ and $z$ directions, the phonon dispersion along $y$ and $z$ directions in the MC simulations should be folded to be compared with that in the AGF calculation. The phonon modal transmissions used in the MC simulation are then obtained by linearly interpolating the corresponding modal transmission calculated by the AGF method. In this work, the phonon modal transmissions used in the MC simulation are referred as phonon modal transmissions by interpolation. For grain boundaries modeled by a-Si interface with different thickness, both the phonon modal transmissions by interpolation and the corresponding phonon modal transmissions by the AGF method are calculated.

## 3. Results

Firstly, the a-Si interface with a thickness of 5 CC (2.716 nm) are used to study the interfacial thermal transport. The phonon modal transmissions across the a-Si interface are calculated by the AGF method. The unit cell of Si contact is set as $1 \times 4 \times 4$ CC, which is the same as that in the Ref.[53]. The phonon modal transmission are plotted in colored dots as shown in Fig. 2 (a), in which the normalized ky = 0.016 and kz = 0.016. The color is scaled by the value of phonon modal transmission according to the color bar. The phonon dispersion are also calculated by GULP[59] and shown by solid line in Fig. 2 (a). Then, the phonon modal transmissions for the MC simulations are obtained by linearly interpolating the corresponding modal transmission calculated by the AGF method. The phonon modal transmissions by interpolation are shown in Fig. 2 (b) by dots according to the color bar, and the corresponding phonon modal transmissions calculated by AGF method are these in Fig. 2 (a). Fig. 2 (c) shows the phonon modal transmissions by interpolation for all the phonon modes (red dots), and the corresponding phonon modal transmissions by the AGF calculations (black dots). As shown in Fig. 2 (c), the phonon modal transmissions are close to unity when the frequency is smaller than 3 THz, and quickly decreases as frequency increases. To investigate how the interface thickness will affect the transmission, the phonon modal transmission for the a-Si interfaces with thickness of 4 CC and 6 CC are also studied. The phonon modal transmissions by interpolation for a-Si interface with thickness of 4 CC (2.172 nm), 5 CC (2.716 nm) and 6 CC (3.259 nm) are shown in Fig. 2 (d). The results show that the phonon transmissions are slightly decreased at low frequency (<3 THz) and gradually decreased at higher frequency (>3 THz) as the thickness of a-Si interface increases.

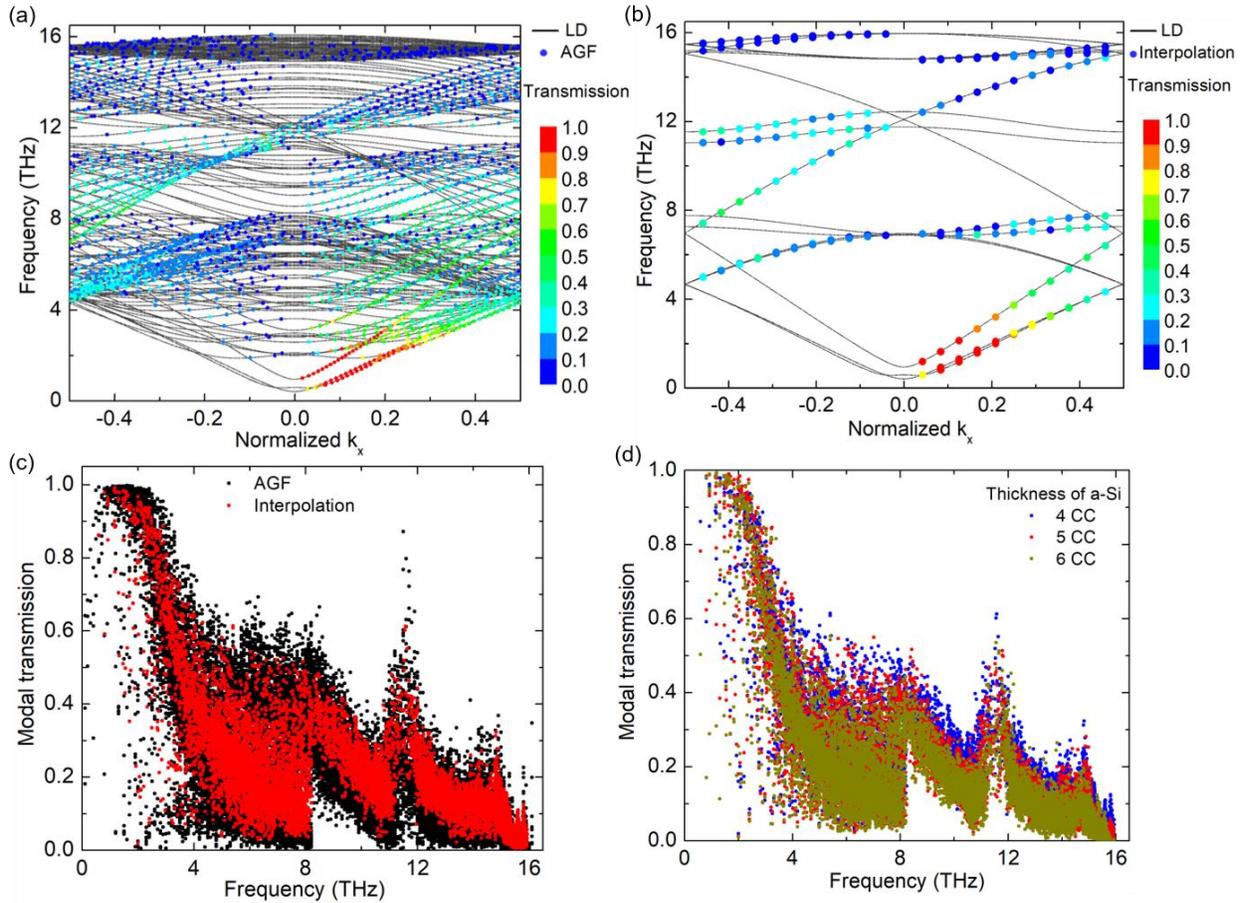

Fig. 2. (a) The phonon modal transmissions calculated by the AGF method (colored dots) for the phonon modes with normalized ky = 0.016 and kz = 0.016. The color represents the value of transmission according to the color bar. (b) Phonon modal transmissions by interpolation for the MC simulations (colored dots), and the corresponding phonon modal transmissions calculated by the AGF method are these in (a). The phonon dispersions (black solid line) in (a) and (b) are calculated by lattice dynamics (LD). (c) Phonon modal transmission by interpolation for all the phonon modes and the corresponding phonon modal transmission calculated by the AGF method. The thickness of a-Si interface is 5 CC (2.716 nm) for the results in (a), (b) and (c). (d) Phonon modal transmissions by interpolation for all the phonon modes across the a-Si interface with thickness of 4 (blue dots), 5 (red dots) and 6 (dark yellow dots) CC.

Based on the phonon modal transmission by interpolation as shown in Fig. 2 (d), the VRMC with full band of phonon dispersion is then applied to investigate the thermal transport in nc-Si (Fig. 1). Fig. 3 (a) shows the ratio of thermal conductivity of nc-Si to that of bulk Si versus the thickness of the a-Si interface. For comparison, the experimental results[16] of the nc-Si with grain sizes of 550 nm (red dashed line) and 76 nm (blue dashed line) are also shown in Fig. 3 (a). The predicted thermal conductivity by MC simulation agrees well with the experiment measurements for the nc-Si with grain sizes of 550 nm and 76 nm. The thickness of the a-Si interface is 5 CC (2.716 nm) and 4 CC (2.172 nm) for nc-Si with grain size of 550 nm and 76 nm, respectively. The modal thermal conductivities of nc-Si and bulk Si are also studied and shown in Fig. 3 (b). The modal $\kappa$ is found to be decreased in the whole frequency range as the grain size decreases. To further understand these results, the distribution of phonon MFPs of bulk Si and nc-Si with grain sizes of 550 nm and 76 nm are compared in Fig. 3 (c). As shown in Fig. 3 (c), the MFPs of low frequency phonons are much larger than the corresponding grain size because of the large phonon modal transmission (Fig.2 (d)). Furthermore, the ratios of MFPs of nc-Si to that of bulk Si are plotted in Fig. 3 (d), which shows that the phonon MFP can be effectively reduced in the whole frequency range as the grain size decreases.

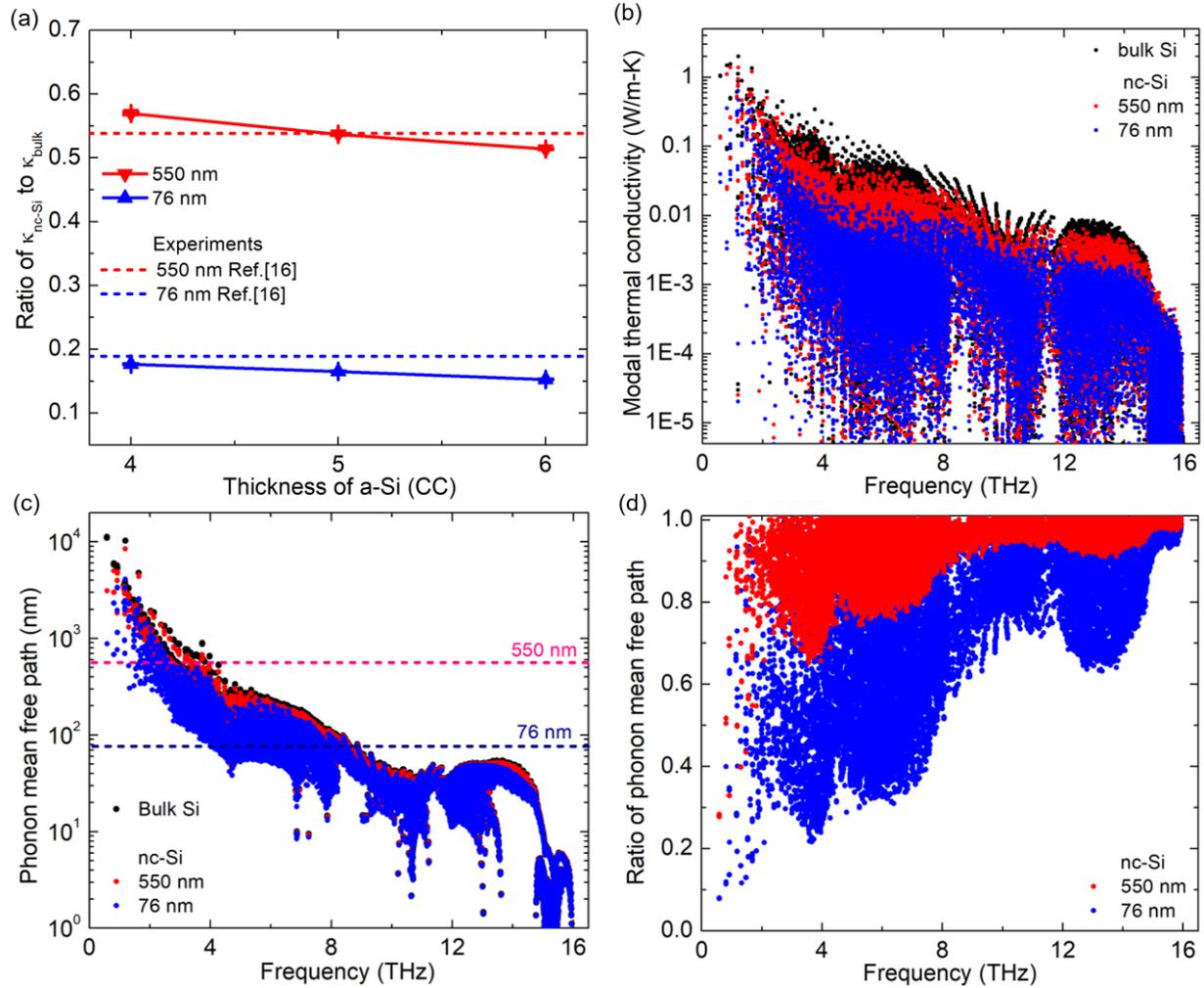

Fig. 3. (a) The ratio of thermal conductivity of nc-Si ($\kappa_{nc-Si}$) with grain sizes of 550 nm (red triangle) and 76 nm (blue triangle) to that of bulk Si ($\kappa_{bulk}$) versus the thickness of a-Si interfaces 4 CC (2.172 nm), 5 CC (2.716 nm) and 6 CC (3.259 nm). The horizontal red and blue dashed lines are from the experimental measurements[16] for the nc-Si with grain size 550 and 76 nm, respectively. (b) The modal thermal conductivity of nc-Si and bulk Si. (c) The phonon mean free path of bulk Si and nc-Si. The pink and blue dashed lines are for reference. (d) The ratios of MFPs of nc-Si to that of bulk Si. In (b) to (d), the nc-Si with grain size 550 nm and 76 nm are represented by the red and blue dots, respectively.

Furthermore, the spectral thermal conductivity of bulk Si and nc-Si with the interface thickness of 4 CC, 5 CC and 6 CC based on the modal κ are shown in Fig. 4 (a). For comparison, the corresponding normalized cumulative thermal conductivities are plotted in Fig. 4 (b). Here, the cumulative thermal conductivities are normalized by the κ of bulk Si. As shown in Fig.4 (a), the contributions of phonon with frequency ranging from 3 to 8 THz in nc-Si is largely reduced because of the reduction of the phonon modal transmission (Fig. 2 (d)). Moreover, as the thickness of a-Si interface increases from 4 CC (2.172 nm) to 6 CC (3.259 nm), the thermal conductivity of low frequency phonons (< 3 THz) is almost unchanged and the overall thermal conductivity is slightly decreased in Fig. 4 (b). To further investigate the grain size effect, the spectral thermal conductivity of nc-Si with grain size 550 nm and 76 nm, and the corresponding normalized cumulative thermal conductivities are shown in Fig. 4 (c) and (d), respectively. The thickness of a-Si interface is 5 CC and 4 CC for the nc-Si with grain sizes of 550 nm and 76 nm, respectively, which can produce the ratio of κ close to the experiment measurements (Fig. 3(a)). The results show that as the grain size decreases, the overall thermal conductivity is largely reduced (Fig. 4 (c)) and the ratio of contribution of low frequency phonons is increased (Fig. 4 (d)), which indicates that grain size has a strong effect on the thermal conductivities in the whole frequency range.

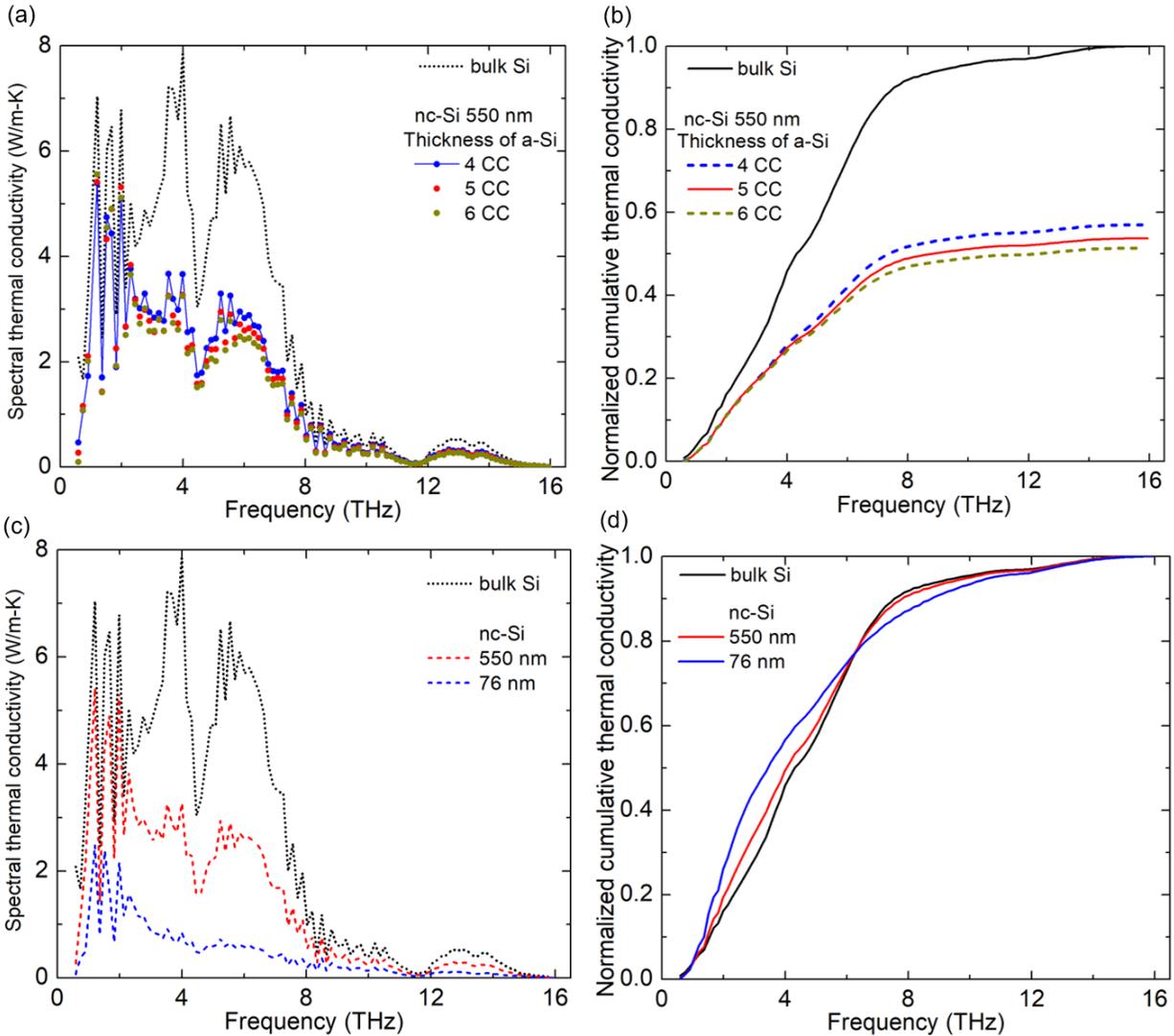

Fig. 4 (a) The spectral thermal conductivity of bulk Si (black dotted line) and nc-Si with grain boundary thickness of 4 CC (blue line dots), 5 CC (red dots) and 6 CC (dark yellow dots) versus the phonon frequency. (b) Normalized cumulative thermal conductivity of bulk Si and nc-Si versus the phonon frequency. The thermal conductivity of nc-Si is normalized by the $\kappa$ of bulk Si. The grain size of nc-Si is 550 nm in (a) and (b). (c) The spectral thermal conductivity of bulk Si (black dotted line) and nc-Si with grain size of 550 nm (red dashed line) and 76 nm (blue dashed line) versus the phonon frequency. (d) The normalized cumulative thermal conductivity of bulk Si and nc-Si with grain sizes of 550 nm (red line) and 76 nm (blue line).

To further investigate the grain size effect, nc-Si with grain sizes varying from 10 nm to 700 nm are studied. The ratio of thermal conductivities of nc-Si to that of bulk Si versus grain size are shown in Fig. 5 (a). For comparison, the experiment measurements of nc-Si with grain sizes 550 nm[16], 144 nm[16], 76 nm [16], 30 nm[60] and 9.7 nm[23] are also plotted in Fig. 5 (a). The results show that the predicted thermal conductivities are quite close to these experimental measurements when the thickness of a-Si interface is around 5 CC (2.716 nm), which implies that the effect of grain boundary on phonon transport can be reasonably modeled by the a-Si interface with 5 CC thickness. The ratio of the thermal conductivity of nc-Si to that of bulk Si is significantly decreased from 54% to 3% as the grain size decreases from 550 nm to 10 nm. The normalized cumulative thermal conductivities versus phonon frequency and phonon mean free path for bulk Si and nc-Si with the grain boundary thickness of 5 CC are shown in Fig. 5 (b) and (c), respectively. These low frequency phonons (<3 THz) transport substantial amounts of heat in nc-Si by contributing 37%, 46%, 50%, 54% and 57% to the total thermal conductivity of nc-Si with the grain sizes of 550 nm, 144 nm, 76 nm, 30 nm and 10 nm, respectively (Fig. 5(b)). Although the grain size can be controlled to a small value, but the phonon MFPs can be much larger than the grain size (Fig.3 (c)). As shown in Fig. 5 (c), the phonons with MFP larger than the corresponding grain size contribute from 30% to 96% in nc-Si as the grain size decreases from 550 nm to 10 nm. These analyses indicate that the VRMC with full band of phonon dispersion using phonon modal transmission by mode-resolved AGF can effectively predict the thermal conductivity of nc-Si close to the experimental results in a wide range of grain size, and the phonon modes with MFP larger than the grain size transport substantial amounts of heat in nc-Si.

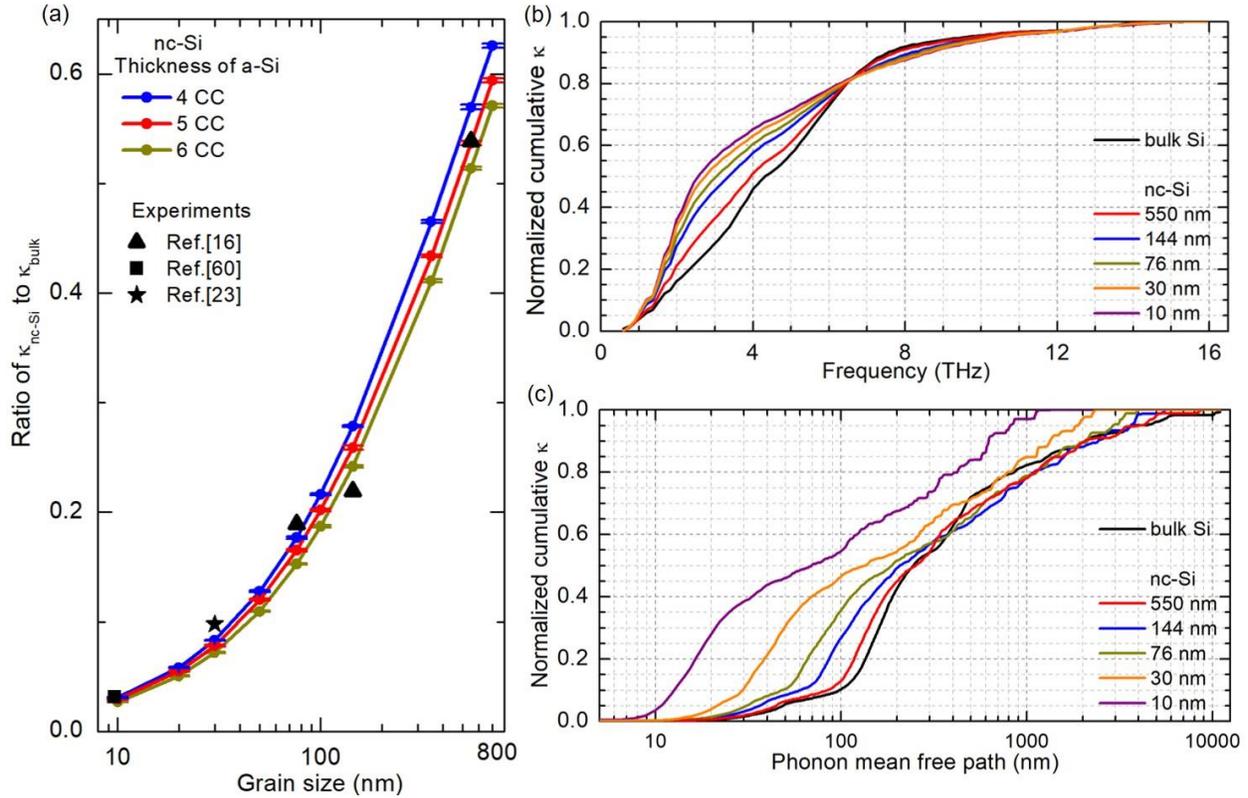

Fig.5. (a) The ratio of the thermal conductivities of nc-Si to that of bulk Si versus grain size. The thickness of interface is varied from 4 CC to 6 CC. The experiment measurements of nc-Si with grain sizes of 550 nm, 144 nm, 76 nm[16] (triangle), 30 nm[60] (square) and 9.7 nm[23] (star) are plotted for comparison. The normalized cumulative thermal conductivity versus frequency (b) and phonon mean free path (c) for bulk Si and nc-Si with grain sizes of 550 nm, 144 nm, 76 nm, 30 nm and 10 nm. The thickness of a-Si interface is 5 CC in (b) and (c).

## 4. Conclusion

In this work, the variance reduced Monte Carlo simulation with full band of phonon dispersion using phonon modal transmission by mode-resolved AGF is applied to study the thermal transport in nc-Si. The grain boundaries are modeled by a-Si interfaces. It is found that the phonon modal transmission across a-Si interface is close to unity at low frequency (<3 THz) and quickly decreases as the frequency increases. The predicted ratio of thermal conductivity of nc-Si to that of bulk Si agrees well with the experimental measurements in a wide range of grain size, which is largely decreased from 54% to 3% as the grain size decreases from 550 to 10 nm. The ratio of contribution of low frequency phonons (< 3 THz) is increased from 37% to 57% as the thickness of interface decreases from 550 nm to 10 nm. The analyses show that reducing the grain size can effectively reduce the phonon MFP in the whole frequency range, while the MFPs of low frequency phonons are much larger than the corresponding grain size because of the large phonon modal transmission. Moreover, the phonons with MFP larger than the corresponding grain size contribute from 30% to 96% in nc-Si as the grain size decreases from 550 nm to 10 nm. As the thickness of a-Si interface increases from 4 CC (2.172 nm) to 6 CC (3.259 nm), the phonon transmission is slightly decreased at low frequency (<3 THz) and gradually decreased at higher frequency (>3 THz), which leads to the small reduction of thermal conductivity of nc-Si. This work shows that the VRMC with full band of phonon distribution using phonon modal transmission by mode-resolved AGF method can be an effective way to study the modal thermal transport in nc-Si, which can provide deep insight into the thermal transport properties of complex nanostructures.

**Acknowledgements**

This work was sponsored by the National Natural Science Foundation of China (Grant No.12004033) and Beijing Institute of Technology Research Fund Program for Young Scholars. (L.Y.)